\documentclass[doublecol]{epl2} 

\title{Non-linear edge dynamics of an Integer Quantum Hall fluid}
\shorttitle{Non-linear edge dynamics of an Integer Quantum Hall fluid}

\author{Alberto Nardin\inst{1} \and Iacopo Carusotto\inst{1}}
\shortauthor{A. Nardin and I. Carusotto}

\institute{                    
  \inst{1} INO-CNR BEC Center and Dipartimento di Fisica, Universit\`a di Trento, I-38123 Trento, Italy.  
}

\abstract{
We report a theoretical study of the linear and nonlinear dynamics of edge excitations of an integer quantum Hall state of non-interacting fermions.
New features beyond the chiral Luttinger liquid picture are anticipated to arise from the interplay of the curvature of the Landau level dispersion and of the Pauli exclusion principle. For long-wavelength perturbations, the microscopic numerical results are captured by a chiral nonlinear hydrodynamic equation including a density-dependent velocity term.
In the wave-breaking regime, shock waves are found to be regularized into a complex ripple pattern by dispersion effects. Our results are of specific relevance for experiments with synthetic quantum matter, in particular ultracold atomic gases.	
}

\begin{document}

\maketitle

\section{Introduction}
The quantum Hall (QH) effect is one of the most surprising and intriguing effects of quantum condensed matter physics. This effect was first discovered in two-dimensional electron gases subject to a strong magnetic field \cite{b.KvK,b.TG}: at sufficiently low temperatures and for suitable integer or rational values of the electron density, the electron gas enters a strongly correlated state characterized by a quantized value of the transverse conductance. Along the lines of the so-called bulk-boundary correspondence, such exotic behaviours can be interpreted in terms of {the} non-trivial topology of the many-body wavefunction in the bulk and of the quantized conductance of chiral states propagating around the edge {system~\cite{Tong}.}

The present work reports a theoretical study of the dynamics of edge excitations, with a special attention to those dispersion and nonlinear effects that go beyond the usual chiral Luttinger liquid
picture of linearly dispersing and non-interacting bosons \cite{b.Wen0,b.Wen1}. 
In order to have an exact microscopic description of the system, we focus on the simplest model displaying the QH effect, namely an {{\em Integer} QH (IQH)} state of spin-polarized neutral fermions in the presence of a strong synthetic magnetic field and of a steep trapping potential. While this model might be an oversimplification for a realistic solid-state system of Coulomb-interacting electrons moving through a disordered potential \cite{b.cooper}, it is the most natural description of ultracold gases of fermionic neutral atoms 
subject to a strong synthetic magnetic field \cite{b.a1,b.a2}, an emerging system for the study of topological states of {matter~\cite{b.a3,Dalibard}.}

As compared to previous works on fractional QH states based on the one-dimensional Calogero model \cite{b.Wiegman1,b.Wiegman2}, our fully two-dimensional theory is able to properly include the main ingredients of the {microscopic} dynamics, namely the curvature of the energy-momentum dispersion of the Landau levels in the trapping potential and the intrinsic nonlinearities due to Pauli exclusion {principle. At the same time, the simplicity of our model allows for an exact numerical solution as well as for perturbative analytical insight in suitable limits.}

The goal of our study will be to {shine light on} the complex nonlinear features displayed by the spatio-temporal dynamics of the density modulation on the {edge in response to classical excitation} potentials of different spatial shapes and different strengths.
For long wavelength perturbations, the numerical results are quantitatively captured by a chiral hydrodynamic description based on a density-dependent propagation {speed. At sufficiently long times wave-breaking effects may set in, but shock waves} get regularized by dispersive terms beyond this simple hydrodynamic description {into large-amplitude ripples.}
These results are a preliminary step in view of attacking the much more challenging case of fractional QH states, for which the nonlinear dynamics is intertwined with the fractional statistics of the excitations \cite{b.Wiegman3}.

\section{The problem and numerical simulations}
\begin{figure}
	\onefigure{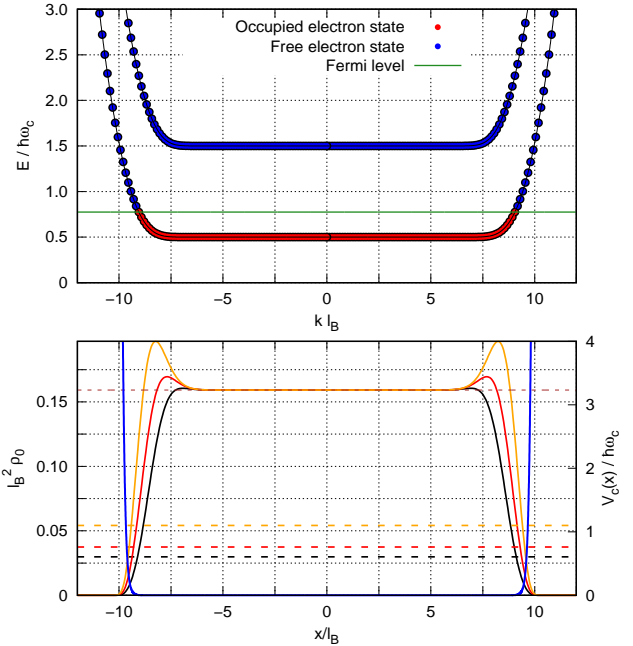}
	\caption{{Top panel: dispersion of the two lowest Landau levels. Dots are coloured in red/blue depending on whether each single-particle is filled/empty in the many-body GS.  			
	Bottom panel: GS density profiles for different Fermi wavevectors $k_F$ (black, red, yellow lines, left $y$ axis); confining potential $V_c(x)$ (blue line, right $y$ axis) and position of the Fermi energy (dashed lines).}
	For clarity, the density of dots in the top panel has been reduced by choosing a smaller $L_y=50\,l_B$ system than in the bottom panel, $L_y=200\,l_B$.}
	\label{fig.0}
\end{figure}

An IQH system {can be described} as an ensemble of {non-interacting spin-polarized fermions} with single particle Hamiltonian~\footnote{In this work we focus on continuous-space geometries, but analogous results have been obtained in Harper-Hofstadter lattices~\cite{b.HH}.}
\begin{equation}
\mathcal{H}_1 = \frac{\boldsymbol{\pi}^2}{2m}+V_c(x)
\end{equation}
where $\boldsymbol{\pi}=\boldsymbol{p}+e\boldsymbol{A}(\boldsymbol{r})$ is the gauge invariant mechanical momentum, $\boldsymbol{A}(\boldsymbol{r})$ is the vector potential and $V_c(\boldsymbol{r})$ a smooth confining potential. The magnetic field $B=\boldsymbol{\nabla}\times\boldsymbol{A}$ is taken as constant.
To simulate a 
{strip geometry with steep edges, periodic boundary conditions $\psi(x,y+L_y)=\psi(x,y)$ are imposed along $y$ and the confinement potential {is} chosen to only depend on $x$. Along this direction}, $V_c(x)$ is taken as steeply rising on {the scale of the magnetic length $l_B=\sqrt{{\hbar}/{eB}}$ and to have a magnitude much larger} than the spacing between different Landau levels {$\hbar\omega_c=\hbar e B /m$, as shown in Fig.\ref{fig.0}}. 

The Landau gauge $\boldsymbol{A}=Bx \hat{y}$ considerably simplifies the problem of finding the {single-particle orbitals in our geometry. Thanks to the translational symmetry along $y$, these are of the form $\psi_{n,k}(x,y)=e^{iky}\,\phi_{n,k}(x)$, where $\phi_{n,k}(x)$ is a solution of the eigenvalue problem}
\begin{equation}
\left(e^{-iky}\mathcal{H}_1e^{iky}\right)\,\phi_{n,k}=E_{n,k}\phi_{n,k}.
\end{equation}
and the wavevector $k$ is quantized to an integer multiple of $2\pi/L_y$.
In the bulk the single-particle orbitals have the form of shifted eigenfunctions of the one-dimensional harmonic oscillator with the (almost) constant energy $(n+1/2)\hbar \omega_c$ of Landau levels. Near the edges {they} get pushed against the steep confining potential $V_c(x)$ and {their} energy rises accordingly forming the chiral edge states. 
The ground state (GS) of the system at zero temperature {is} built by filling all the states below the Fermi {energy.}
{In the following, the Fermi energy is chosen to be located in between} the lowest and the first excited Landau level {so to focus on a single chiral edge channel} as shown in the upper panel of Fig.\ref{fig.0}. We indicate with $v$ the Fermi velocity at the Fermi point $k_F$ separating the regions of filled and empty states.
The {edge dynamics is then} probed by applying an external time-dependent perturbation potential $V_e(\boldsymbol{r},t)$ onto the system. {In the absence of interparticle interactions, every single-particle orbital then evolves in time independently from the others and the only correlations are the ones stemming from the Pauli principle.} 
{For analytical simplicity we adopt an excitation potential $V_e(y,t)$ that only depends on $y$ and $t$, but we have verified that perturbations localized on the edge would give qualitatively similar results for the late time dynamics after the excitation potential has been switched off. The external potential is assumed to be} turned on and then off {on a time-scale $\tau$ that is slow compared to the bulk dynamics ($\tau\gg \omega_c^{-1}$),} but fast enough to excite the chiral edge modes of frequency $vq$. Throughout this work, the characteristic wavevector $q$ of the excitation potential is assumed to be much smaller than the inverse radius of the quantum cyclotron orbits.

{The key observable of our work is the system density $\rho(\mathbf{r})$. This is obtained as the diagonal part of the one-body density matrix, which for non-interacting particles is given by $\rho(\mathbf{r}',\mathbf{r}, t)=\sum_\alpha \psi^*_\alpha(\mathbf{r}', t) \psi_\alpha(\mathbf{r}, t)$. }
The GS density $\rho_0(x)$ is plotted for different {values of the Fermi energy (and thus of the Fermi momentum $k_F$)} in the bottom panel of Fig.\ref{fig.0}.
{Under the assumed $\tau\gg \omega_c$ condition, the excitation potential can not induce
transitions to excited Landau levels and the bulk density remains equal to its incompressible value $\rho={\nu}/{(2\pi l_B^2)}$ for integer-valued filling $\nu$~\cite{Tong}, for our parameters equal to $\nu=1$. The density variation $\delta\rho=\rho-\rho_0$ will thus be}  significatively different from zero only within a few magnetic lengths from the sharp system boundary{. 
In what follows, we will focus on an effective one-dimensional description of the edge, obtained by integrating the two-dimensional density profile along the orthogonal direction over half a sample, $\delta\rho_e(y)=\int_0^\infty \delta\rho(x,y) dx$}

If the external excitation is {slowly varying in space}, the {dispersion of edge modes can be linearized around the Fermi momentum. {For large enough systems, left and right edge channels are decoupled and} the effective density variations $\delta\rho_e(y)$ on both edges obey linear chiral hydrodynamic equations \cite{b.a9} with the Fermi velocity $\pm v$ and a source term proportional to the spatial gradient of the external potential $U_e$,}
\begin{equation}
\label{eq:linear_dynamics}
\partial_t\delta\rho_e=\pm v \,\partial_y \delta\rho_e - \frac{1}{2\pi\,\hbar}\partial_y U_e.
\end{equation}
The potential $U_e(y,t)=\int V_e(x,y; t) |\phi_{n,k_F}(x)|^2 \,dx$ {is the effective potential experienced by the edge orbitals}
and the plus/minus signs indicate chiral propagation towards negative/positive values of the $y$ coordinate on the right/left edges. The form of the source term in Eq.\ref{eq:linear_dynamics} corresponds to the transverse Hall current induced by the force $-\partial_y U_e$ that directly depletes or replenishes the density on the edge. This is radically different from the one appearing in the case of a one-dimensional classical gas, where the density modulation is instead related to the gradient of the force, that is the second spatial derivative of the potential. Mathematically it can be motivated by bosonizing the quadratic interaction in second quantization and integrating out the transverse direction. According to Eq.\ref{eq:linear_dynamics}, once the external excitation has been turned off, {the density modulation} $\delta\rho_e$ rigidly {propagates} at the Fermi velocity $v$. 

{To go beyond this chiral Luttinger liquid approach~\cite{b.Wen0,b.Wen1}, an exact numerical study of the dynamics of each single-electron orbital is performed. In our calculations, a separable form of the external potential is used, 
$V_e(y,t)=\lambda \,g(y) \exp\left[-(t-t_0)^2/\tau^2\right]$ with 
a Gaussian temporal profile of duration $\tau$ and centered at $t_0\gg\tau$. In what follows, different forms of $g(y)$ are considered to highlight different features of the dynamics.}



\begin{figure}
	\onefigure{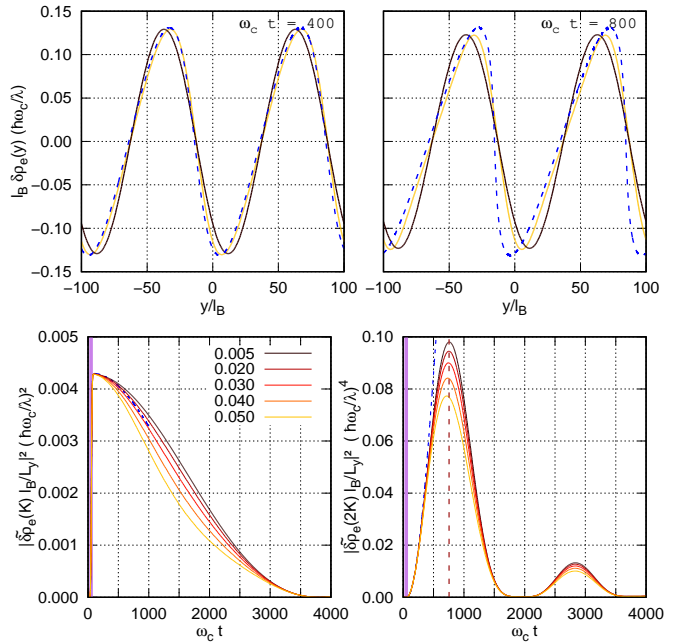}
	\caption{Upper panels: snapshots at different times of the {density modulation $\delta\rho_e(y)$ generated on the negative-$x$ edge by a spatially periodic excitation of wavevector $K$. Different curves are for growing values of the excitation potential strength $\lambda$ (black to red to yellow). 
	Bottom panels: time-dependence of the Fourier components of $\delta\rho_e(y)$ at $K$ (left) and at its second harmonic $2K$ (right) for the same excitation strengths. 
	Purple shaded regions indicate the excitation transient.
	Dashed blue curves are the solution of eq.\ref{eq:NonLinear1} for $\lambda/\hbar\omega_c=0.05$. In the bottom panels these are plotted approximatively up to the wave-breaking instant.
	System parameters: $L_y=400l_B$ and $k_F\simeq 8.58l_B^{-1}$, corresponding to $N=1093$ fermions and a Fermi energy $E_F\simeq0.61\hbar\omega_c$. Fermi velocity $v\simeq0.25 l_B\omega_c$ and curvature $c\simeq 0.42l_B^2\omega_c>0$. Excitation wavevector $K=4 \times {2\pi}/{L_y}$, centred in time at $t_0=50\omega_c^{-1}$ with a width $\tau=15\omega_c^{-1}$.}}
	\label{fig.1}
\end{figure}

\section{Extended sinusoidal excitation}
\begin{figure}
	\onefigure{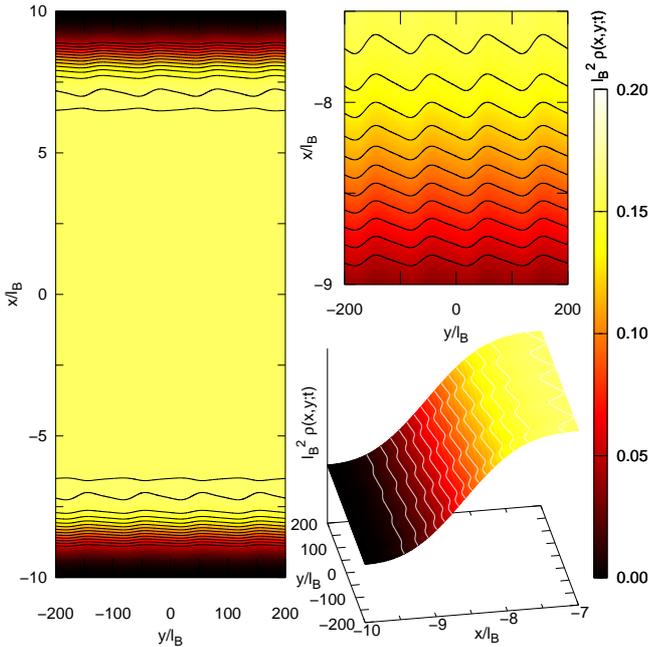}
	\caption{{Left panel: heat-map snapshot at $t=1000/\omega_c$ of the modulated system density} $\rho(x,y)$ in response to a strong and {spatially} periodic excitation. Top-right panel: zoom on a small portion of the system on the negative-$x$ edge. Bottom-right panel: same data displayed as a surface plot. 
	Same system parameters as for the strongest $\lambda=0.05\hbar\omega_c$ curves of Fig.\ref{fig.1}.}
	\label{fig.3}
\end{figure}

As a first step, we consider the simplest case of a spatially periodic excitation potential with $g(y)=\sin^2\left(Ky/2\right)$. This leads to a correspondingly periodic density modulation $\delta\rho_e(y,t)$ which, on the {$x<0$ edge, propagates in the positive-$y$ direction.
In the upper panels of Fig.\ref{fig.1}, we show {two snapshots of $\delta\rho_e(y,t)$}  in the linear regime of a weak excitation potential (black line) and we compare them to the same curve in a stronger excitation regime where the non-linearity is relevant (yellow line). Here,} the sinusoidal wave of the linear response {deforms into} a sawtooth pattern, with the compression regions moving faster and the decompression ones slower, effectively producing a sharp front edge and a smoother trailing one. A {complete plot of the density profile in the whole system} is shown in Fig.\ref{fig.3} for the same configuration: {as expected,} the bulk is {not affected by the external potential $V_e$} and only the edges get excited. The sawtooth-shaped deformation due to the nonlinearity is clearly visible on the iso-density lines that are displayed in all panels. 

This nonlinear behaviour can be heuristically explained as follows. As usual for degenerate Fermi gases, the Fermi wavevector $k_F$ can be related to the average number of electrons per unit length in the $y$ direction. Restricting to the half-stripe $x<0$, we have
$\rho={N_e}/{L_y}=
{k_F}/{(2\pi)}$.
An increase in the particle number density from $\rho$ to $\rho+\delta\rho$ then corresponds to a shift of the Fermi wavevector by 
$\Delta k_F=2\pi\,\delta\rho$.
Because of the curvature of the dispersion, the shift in $k_F$ implies a corresponding change in the Fermi velocity, which at the lowest order reads
\begin{equation}
\label{eq:density_velocity_shift}
v'=v+c\Delta k=v+2\pi c\, \delta\rho.
\end{equation}
For the confinement potentials under consideration here, the curvature is positive $c>0$ as shown in Fig.\ref{fig.0}. Within a local density approximation, we can consider the (local) Fermi velocity to be increased in the compressed regions of the Fermi gas, and vice-versa to be decreased in the rarefied regions.
Based on these heuristic arguments, the chiral hydrodynamic Eq.\ref{eq:linear_dynamics} may then be generalized to a non-linear hydrodynamic equation
\begin{equation}
\label{eq:NonLinear1}
\partial_t\delta\rho_e=\pm (v + 2\pi c \, \delta\rho_e)\,\partial_y \delta\rho_e - \frac{1}{2\pi\,\hbar}\partial_y U_e
\end{equation}
which is expected to hold for long-wavelength excitations. 

Eq.\ref{eq:NonLinear1} has an implicit solution due to Riemann $\delta\rho_e=F\left(y\pm(v+2\pi c\,\delta\rho_e)t\right)$, with $F$ an arbitrary function. At not too large times this analytical solution perfectly captures the steepening of the sinusoidal modulation and its deformation into a sawtooth profile, as shown by comparing the solid and dashed lines in the upper panels of Fig.\ref{fig.1}. 
At later (yet finite) times, however, it predicts overturns, that is multivalued unphysical solutions~\cite{b.a10}. 
As it is shown by our numerics in the following figures, this pathological behaviour of the analytical approximation gets regularized in the complete theory, where the density profiles remain smooth at all times. 
 
Further light on the dynamics of the system is offered in the bottom panels of Fig.\ref{fig.1}, which show the time-dependence of the square moduli of the Fourier components of the density $\widetilde{\rho}_e(q)=\int \rho_e(q)\,e^{-iqy}dy$ for the fundamental and harmonic modes at $q=K,2K$ (left, right) and different external potential strengths (black to yellow lines).
Dispersive effects due to the finite curvature $c>0$ are responsible for a decay of the excitation at late times (left panel), even in the linear regime. 
The mechanism underlying this decay can be located in the interference between the single-particle orbitals involved in the excitation at wavevector $q$, that span a wavevector region from $k_F-q$ to $k_F$. As such, the decay of weak excitations can not be accounted for by the hydrodynamic Eq.\ref{eq:NonLinear1}. Analytical insight into it will be offered in the next section by a microscopic perturbation theory on the single-particle orbitals.
The contribution to the decay of the fundamental mode at $q=K$ due to up-conversion processes to the mode at $q=2K$ by the non-linearity of eq.\ref{eq:NonLinear1} is significant for stronger excitations and is visible as a dashed line in the bottom-left panel of Fig.\ref{fig.3}.

The bottom-right panel shows instead the time-evolution of the second harmonic of the density modulation at $q=2K$. This is generated by nonlinear effects and, at moderate excitation strengths, scales as the square of the fundamental excitation at $q=K$. Interestingly, the linear growth of the harmonic component at early times is well captured by the hydrodynamic Eq.\ref{eq:NonLinear1}, as it is shown by the dashed line. The later dynamics is instead dominated by single-particle interference effects. The superposition of the fundamental and harmonic components is responsible for the sawtooth deformation of $\delta\rho_e(y,t)$.

\section{Time-dependent perturbation theory}
An alternative strategy to get analytical insight in the decay of the excitations and in the non-linear response found numerically is based on a first and second order perturbation theory for the time-evolution of the single-particle orbitals. Thanks to the simple analytical form of the excitation, closed-form results can be obtained for these quantities. 

At lowest order in $\lambda$, only the fundamental density component $\widetilde{\rho}_e(K)$ at the external potential wavevector is non-zero; in the thermodynamic limit $L_y\rightarrow\infty$ we obtain that it decays in time proportionally to $\mbox{sinc}\left[{c}K^2(t-t_0)/2\right]$, yielding a finite lifetime on the order of $T_d={\pi}/{c K^2}$.
The sinc-shaped behaviour is caused by the sharp discontinuity of the fermionic occupation at the Fermi point at $k_F$, which acts as an ideal low-pass filter cutting all the frequencies beyond $vK+{cK^2}/{2}$: at linear perturbative order no fermion can get excited beyond $k_F+K$ and the highest frequency contained in $\delta\rho_e$ comes from the $k_F\to k_F+K$ transition. Of course, any finite temperatures will smear the Fermi edge, giving a faster time-decay.
A natural question for follow-up work is whether the decay persists in the presence of (weak) interactions, e.g. p-wave interactions between spin-polarized fermions and, more importantly, whether the Fermi edge is stable against such interactions. For Coulomb interactions, this question was addressed in~\cite{b.a11} finding interesting edge reconstruction effects.

Using next-to-leading order perturbation theory, we are also able to obtain an approximate form for the harmonic density modulation, 
 $\widetilde{\rho}_e(2K)\propto{\sin^2\left[cK^2(t-t_0)\right]}/{(cK^2(t-t_0))}.$
This formula gives a characteristic timescale for the onset of lowest-order non-linear phenomena as $T_{NL} \propto 1/cK^2$ (shown as a dashed brown vertical line in the bottom-right panel of Fig.\ref{fig.1}.), 
which by no coincidence is of the same order of magnitude of the decay time $T_d$.
As expected, in the $c\rightarrow0$ limit, the fundamental component $\widetilde{\rho}_e(K)$ approaches the solution of Eq.\ref{eq:linear_dynamics} while the harmonic one vanishes $\widetilde{\rho}_e(2K)\rightarrow0$, showing that the linear theory is exactly recovered in the limit of a flat edge dispersion.
Finally, it is interesting to note that the corrections to $\widetilde{\rho}_e(K)$ (dispersive effects) are $\mathcal{O}(c^2)$, while those to $\widetilde{\rho}_e(2K)$ (non-linear effects) are $\mathcal{O}(c)$: for small curvature parameters, far away from a shock region, dispersive effects will be of higher order when compared to the non-linear ones\footnote{No other curvature factors are present in the formulae.}.

\begin{figure}
	\onefigure{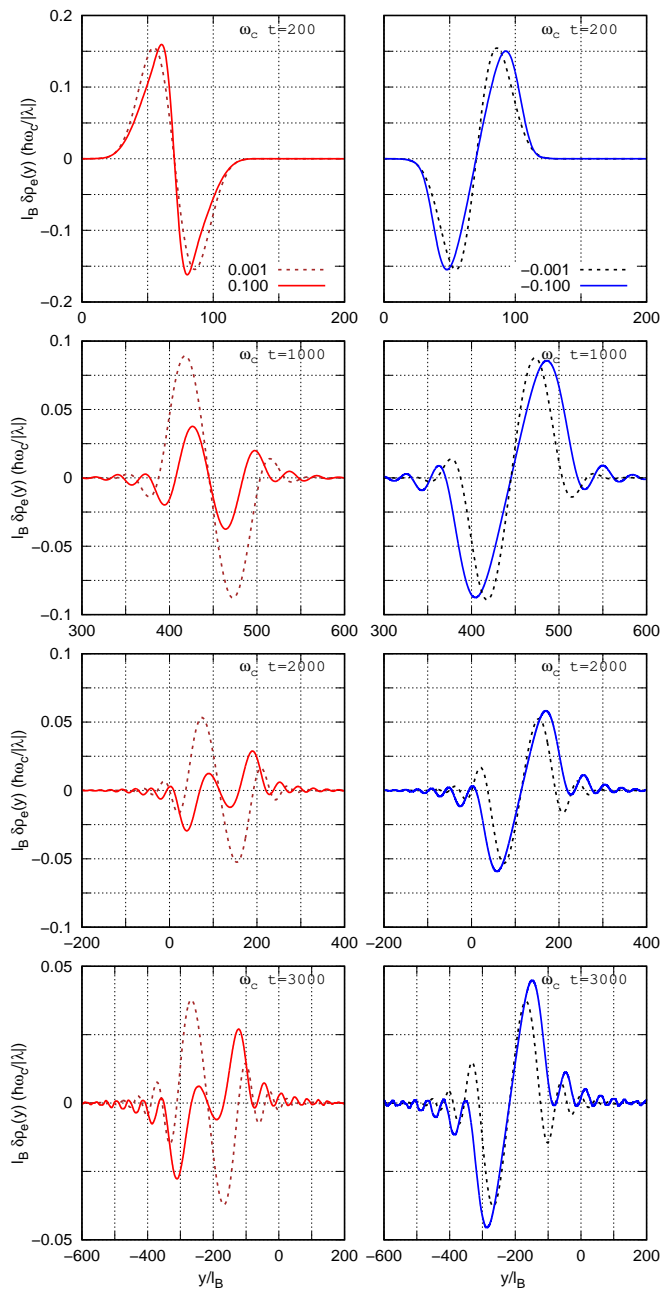}
	\caption{From top to bottom: snapshots at growing $t$  of the density modulation $\delta\rho_e(y)$ on the negative-$x$ edge in response to a Gaussian-shaped excitation with $\lambda>0$ (left) and $\lambda<0$ (right). Within each panel, solid and dashed lines refer to strong and weak excitations with $|\lambda|/\hbar\omega_c=0.001$ and $0.1$. System parameters: $L_y/l_B=800$ and $k_F\simeq 9.02l_B^{-1}$, corresponding to $N=2297$ fermions below $E_F\simeq0.76\hbar\omega_c$, $v\simeq0.47 l_B\omega_c$, and $c\simeq 0.57l_B^2\omega_c$. Gaussian excitation with $\sigma/l_B=20$ centered at $t_0=50\omega_c^{-1}$ of width $\tau=15\omega_c^{-1}$.}
	\label{fig.2}
\end{figure}

\section{Gaussian excitation}
Based on the understanding accumulated on spatially periodic excitations, we can now proceed to consider the response to spatially localised excitations of different forms. We start from a Gaussian-shaped $g(y)=\exp\left(-{y^2}/{\sigma^2}\right)$ that, in agrement with Eq.\ref{eq:linear_dynamics}, produces at short times an anti-symmetric two-lobed density modulation.
At linear regime (dashed lines), the deformation that is visible at later times is due to the same interference effects that were responsible for the decay of the spatially periodic modulation discussed above and swaps signs under a change of the sign of $\lambda$.

This symmetry is no longer valid for stronger excitations (solid lines). In this case, the distortion is much stronger for $\lambda>0$ (left column) than for $\lambda<0$ (right column). This behaviour can again be qualitatively understood in terms of the nonlinear terms in Eq.\ref{eq:density_velocity_shift}: if a compressed region with $\delta\rho_e>0$ is located behind a rarefied region with $\delta\rho_e<0$, the effect of the positive curvature $c>0$ will be to push the two regions against each other. As a result, a shock wave will eventually form between the two, giving rise to large ripples by dispersive effects. In the opposite case, the two regions tend to separate, leaving a smooth transition in between. In this case, some dispersive ripples will of course appear on the outer parts of the density pattern, but have a weaker magnitude.

\section{Sigmoid excitation}
\begin{figure}
	\onefigure{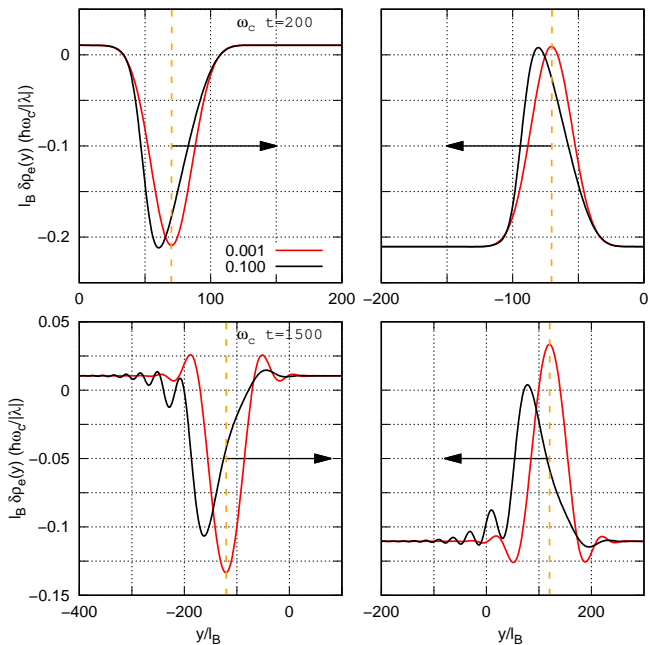}
	\caption{Left column: snapshots at growing $t$ (top to bottom) of the density modulation $\delta\rho_e(y)$ on the negative-$x$ edge in response to a sigmoid-shaped excitation with $\sigma=20\,l_B$. Red (black) curves refer to weak (strong) excitations $\lambda/\hbar\omega_c=0.001$ ($0.1$). Right column, same plots on the positive-$x$ edge propagating in the opposite direction along $y$. All other parameters as in Fig.\ref{fig.2}. Vertical orange dashed lines correspond to the central position of the wavepacket moving at the group velocity, $y=\pm v(t-t_0)$.}
	\label{fig.4}
\end{figure}

In order to produce an initial bell-shaped density excitation, a sigmoid-shaped excitation of the form
$g(y)=\mbox{erf}\left({y}/{\sigma}\right)$ can be used. The results~\footnote{Note that in this case a linear potential must be added to satisfy periodic boundary conditions along $y$ but this is irrelevant in the thermodynamic limit.}  are shown in Fig.\ref{fig.4}. For strong (black lines) and negative perturbations, the trailing edge of the wavepacket gets steeper during propagation because of nonlinear effects (upper-left) and eventually develops density ripples (lower-left). For an equally strong but positive perturbation, the shock-wave behaviour occurs on the leading edge.

Some interesting dynamics is also visible for weak excitations in the linear regime (red lines). Due to the wavevector dependence of the mode decay time $T_d\sim 1/K^{2}$ discussed above, a localised density packet will not only decay but also spread in real space as $\sqrt{c(t-t_0)}$ \cite{b.a12}. As in the previous Fig.\ref{fig.2}, weak oscillations also appear around the main wavepacket as an additional consequence of interference between the different single-particle orbitals participating to the excitation.

\section{Spatially localized V-shaped excitation}
\begin{figure}
	\onefigure{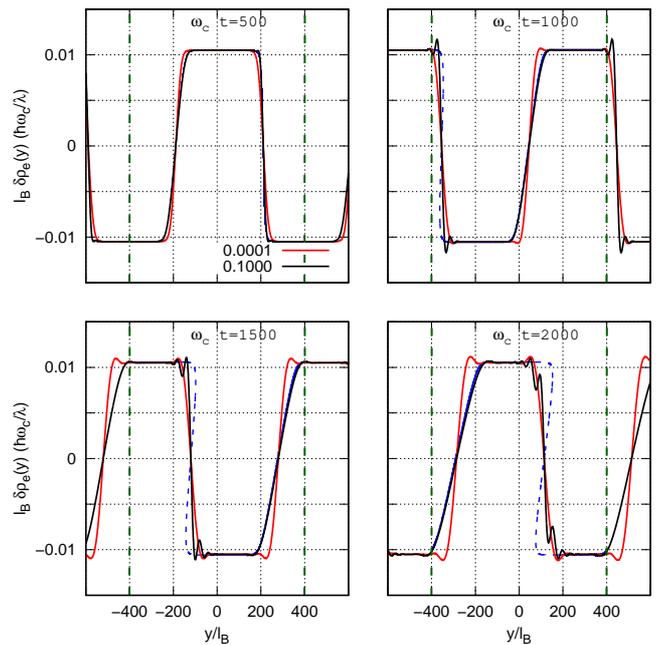}
	\caption{Snapshots at growing $t$ of the density modulation $\delta\rho_e(y)$ on the negative-$x$ edge in response to a V-shaped excitation with $\sigma=20l_B$. Red (black) curves refer to weak (strong) excitations $\lambda/\hbar\omega_c=0.0001$ ($0.1$). Blue curves are the solutions of Eq.\ref{eq:NonLinear1} for $\lambda/\hbar\omega_c=0.1$, where the overturns are indicated by the dashed line. All other parameters as in Fig.\ref{fig.2}. Vertical dark-green dashed lines denote the system boundaries. Beyond these lines, the solutions are extended according to the periodic boundary conditions for clarity.}
	\label{fig.5}
\end{figure}

As a last example, a steep square-wave density modulation has been produced using a suitable V-shaped excitation $g(y)\simeq \ln\left[2\cosh\left({y}/{\sigma}\right)\right]$ 
The results are shown in Fig.\ref{fig.5}. Once again, the non-linearities make the evolution of trailing edges of the square wave to quite differ from that of the front one. The latter remains steep since the fast high density region located behind is pushed against the slower low density region in front of it. 
Eventually, this leads to a shock wave in the front edge that is regularized into ripples, in stark contrast with the overturns displayed by the solution of Eq.\ref{eq:NonLinear1} (blue line).
The trailing edge gets instead smoother and does not display any ripple. The excellent agreement with the solution of Eq.\ref{eq:NonLinear1} in this region confirms that dispersive effects are minimal here and, in particular, no shock occurs. 
\begin{figure}
	\onefigure{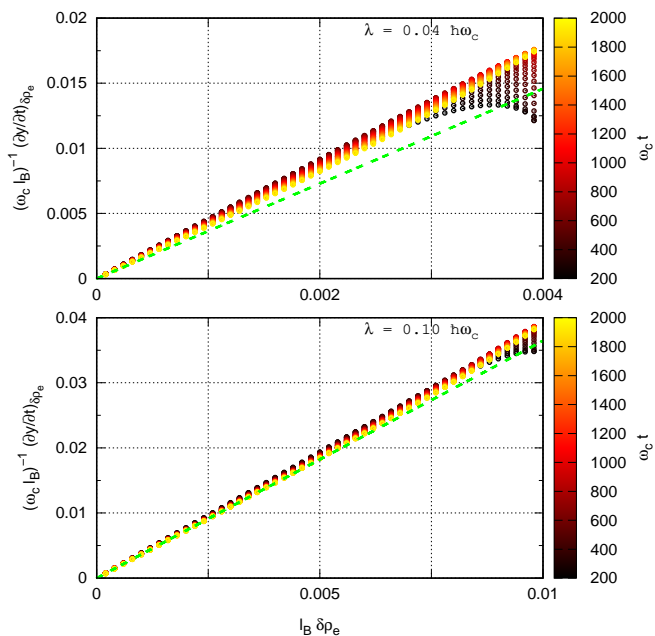}
	\caption{
	We compare the numerically calculated value of the derivative $\frac{\partial y'}{\partial t'}$ at given $\delta\rho_e$ (colored points) with the result Eq.\ref{eq:nl_broadening} obtained from Eq.\ref{eq:NonLinear1} as a function of $\delta\rho_e$ (green dashed line). The color (black to yellow) of the points indicates the different $t$. Upper (lower) panel is for weak (strong) excitation $\lambda/\hbar\omega_c=0.04$ ($0.1$). Same parameters as in Fig.\ref{fig.5}.}  
	\label{fig.6}
\end{figure}
This form of excitation is thus ideal to quantitatively assess the accuracy of the analytical approximation in Eq.\ref{eq:NonLinear1}. This predicts that
\begin{equation}
\label{eq:nl_broadening}
\left(\frac{\partial y'}{\partial t'}\right)_{\delta\rho_e}=2\pi c\, \delta \rho_e
\end{equation}
where the left hand side derivative is taken at a fixed $\delta\rho_e$ and $y'=y-vt$, $t'=t$ are comoving frame coordinates. This relation has been numerically checked for the trailing edge where dispersive phenomena are less relevant. The numerical results are shown in Fig.\ref{fig.6}. As $\lambda$ increases, the overall agreement with Eq.\ref{eq:nl_broadening} gets better.


\begin{figure}
	\onefigure{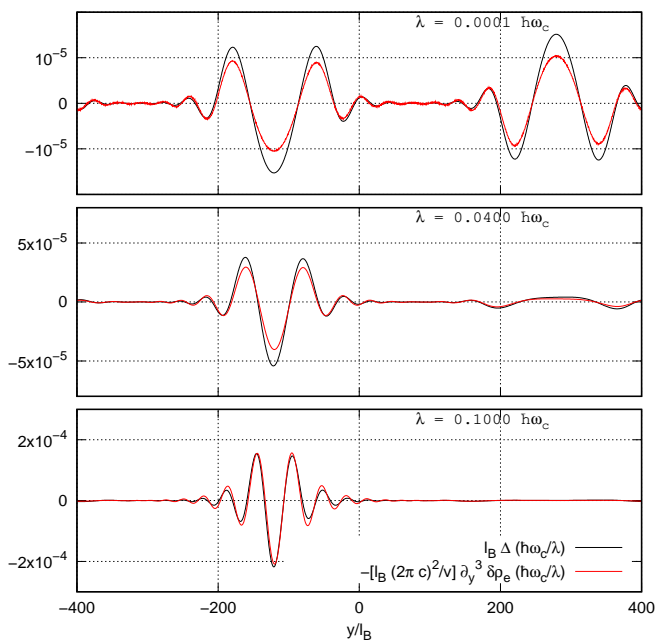}
	\caption{Plot of the suitably normalized difference $\Delta = \partial_t\delta\rho_e + (v+2\pi c\,\delta\rho_e)\partial_y\delta\rho_e$ for a V-shaped excitation (black) localized on the negative-$x$ system edge compared to the suitably normalized third-order derivative term $\partial_y^3\delta\rho_e$ at time $\omega_c t=1500$ and for different excitation strengths (top to bottom).}
	\label{fig.7}
\end{figure}

The physical origin of the deviation is investigated in Fig.\ref{fig.7}, where we plot the spatial profile of the difference $\Delta = \partial_t\delta\rho_e + (v+2\pi c\,\delta\rho_e)\partial_y\delta\rho_e$ (the positive sign is used since we are focusing on the negative-$x$ sample edge). This is localised in regions where the density is most quickly varying. In agreement with the evolution of the waveform discussed above, for strong excitations the dispersive corrections are much larger for the steeper and possibly oscillating front edge rather than for the smoother trailing one.
But most remarkably, comparison of the black and red curves shows how $\Delta$ is accurately reproduced by (minus) the third spatial derivative of the edge density $\delta\rho_e$, in particular for strong excitations. In doing the comparison, the third derivative was heuristically scaled by $v^{-1}(2\pi c)^2$ on the basis of dimensional arguments and arguing that the underlying microscopic time and length scales should not matter in this excitation regime. 
On this basis, we conclude that the first nonlinear correction to Eq.\ref{eq:NonLinear1} for strong modulations must have a form proportional to the third derivative $\partial_x^3\delta\rho_e$, giving an effective Korteweg-de Vries dynamics: the behaviour observed in our numerics is indeed resemblant of shock-waves emerging from such an equation \cite{b.pavloff}. While this conclusion is very accurate for strong excitations (lower panel of Fig.\ref{fig.7}), important corrections are still present in the weak excitation regime (upper panel): the KdV corrections are in fact not able to capture those interference effect that originate from the sharp Fermi edge and that are responsible for the linear damping and spreading of wavepackets.

\section{Conclusions}
In this work we have reported a microscopically exact calculation of the linear and nonlinear edge dynamics of an IQH system of non interacting fermions. For weak perturbations of the GS, a perturbative description is able to capture the effect of the curvature of the single particle dispersion on the propagation of the excitation wavepackets. For stronger perturbations, the edge hydrodynamics displays important non-linear features; in the wave-breaking regime, dispersive effects regularize the shock wave into large-amplitude ripples similarly to the KdV equation.
While a great deal of the non-linear effects can be included in a one-dimensional, chiral hydrodynamic equation for the classical density, a future task will be to understand the origin of the damping, and whether it can be included in the hydrodynamic semiclassical description through dispersive, non-local and/or quantum fluctuations terms.
Another exciting question to be addressed is whether our non-interacting system can support solitonic solutions. The long-term perspective is to extend our microscopic approach to fractional quantum Hall states~\cite{Tong}, where the anyonic statistics of excitations is anticipated to interplay with the nonlinear dynamics to give, for instance, fractional solitons \cite{b.Wiegman3}.

\acknowledgments
We acknowledge financial support from the European Union via the FET-Open grant “MIR-BOSE” (n. 737017) and the H2020-FETFLAG-2018-2020 project “PhoQuS” (n.820392), from the Provincia Autonoma di Trento, partly through the Q@TN initiative, and  from Google via
the quantum NISQ award. IC is indebted to Nicolas Pavloff for stimulating discussions that triggered this research.

\end{document}